\documentclass{article}
\usepackage{spconf,amsmath,graphicx}
\usepackage{hyperref}

\usepackage{xcolor} 
\usepackage{url}

\title{10 hours data is all you need}

\begin{document}

\name{Zeping Min$^{\star}$ \qquad 
Qian Ge$^{\star}$ 
\thanks{$^{\star}$Equal contribution} \qquad 
Zhong Li$^{\dagger}$
\thanks{$^{\dagger}$Corresponding author}}
\address{$^{\star}$ Peking University \\
$^{\dagger}$ Microsoft Research Asia}

%
\maketitle
\begin{abstract}
We propose a novel procedure to generate pseudo mandarin speech data named as CAMP (\underline{c}haracter \underline{a}udio \underline{m}ix u\underline{p}), which aims at generating audio from a character scale. We also raise a method for building a mandarin character scale audio database adaptive to CAMP named as META-AUDIO, which makes full use of audio data and can greatly increase the data diversity of the database. Experiments show that our CAMP method is simple and quite effective. For example, we train models with  10 hours of audio data in AISHELL-1 and pseudo audio data generated by CAMP, and achieve a competitive 11.07 character error rate (CER). Besides, we also perform training with only 10 hours of audio data in AIDATATANG dataset and pseudo audio data generated by CAMP, which again achieves a competitive 8.26 CER.
\end{abstract}

\section{Introduction}
\label{sec:intro}
Training a practical neural network (NN) model often requires a large amount of labeled data. However, obtaining annotated data in application is usually rather expensive and labor-intensive. There are lots of efforts to reduce the dependence of NN models on huge amounts of data, such as \cite{zhang2017mixup}, and \cite{lee2013pseudo}. In the field of speech recognition, it is equally necessary to provide sufficient training data for deep NN models. Inspired by the pseudo label method (\cite{lee2013pseudo}) and mix up method (\cite{zhang2017mixup}), we propose a novel procedure to generate pseudo-labeled data, named as character audio mix up (CAMP), to alleviate the heavy data dependence in automatic speech recognition. 

In summary, our contributions are as follows:
\begin{itemize}
    \item We successfully combine the advantages of pseudo label semi-supervised learning and mix up data augmentation to propose a novel, simple and effective procedure to generate pseudo-labeled speech data named as character audio mix up (CAMP).
    \item We propose a META-AUDIO method for building a mandarin character scale audio database adaptive to the CAMP. The META-AUDIO takes full advantage of audio data and can greatly increase the data diversity in the database, as well as reduce the difficulty of building the database.
    \item Experiments show that the CAMP and META-AUDIO method are simple but quite effective. Training models with 10 hours of audio data in AISHELL-1 together with pseudo audio data generated by CAMP, we achieve a competitive 11.07 character error rate (CER). Besides, we also perform training with only 10 hours of audio data in the AIDATATANG dataset and pseudo audio data generated by CAMP, and again achieves a competitive 8.26 CER.
\end{itemize}

\section{Related work}
A lot of effort has been made to  obtain satisfying modeling given a limited size of training samples. In application, one of the most effective ways is data augmentation \cite{salamon2017deep}, and \cite{park2019specaugment}
, which is often designed carefully based on the nature of data itself, and hence implies a certain pertinence. For instance, in the field of computer vision (CV), common approaches of data augmentation  (\cite{shorten2019survey}) typically include cropping, rotation, mix up (\cite{zhang2017mixup}, \cite{inoue2018data}) and so on, which are specifically developed for images. In the field of automatic speech recognition, the data augmentation is often conducted as follows. One way is to perform data augmentation from the frequency domain. For example,  \cite{jaitly2013vocal} implemented data augmentation by a random linear transformation on the frequency dimension of the spectrogram. Another way to perform data augmentation is from the time domain. For example, in \cite{hannun2014deep}, a large amount of audio in a noisy environment was synthesized by mixing the clean audio with the noises, followed by an appropriate filtering through the average power.

\begin{figure*}
    \centering
    \includegraphics[width=.95\textwidth, height=0.35\textwidth]{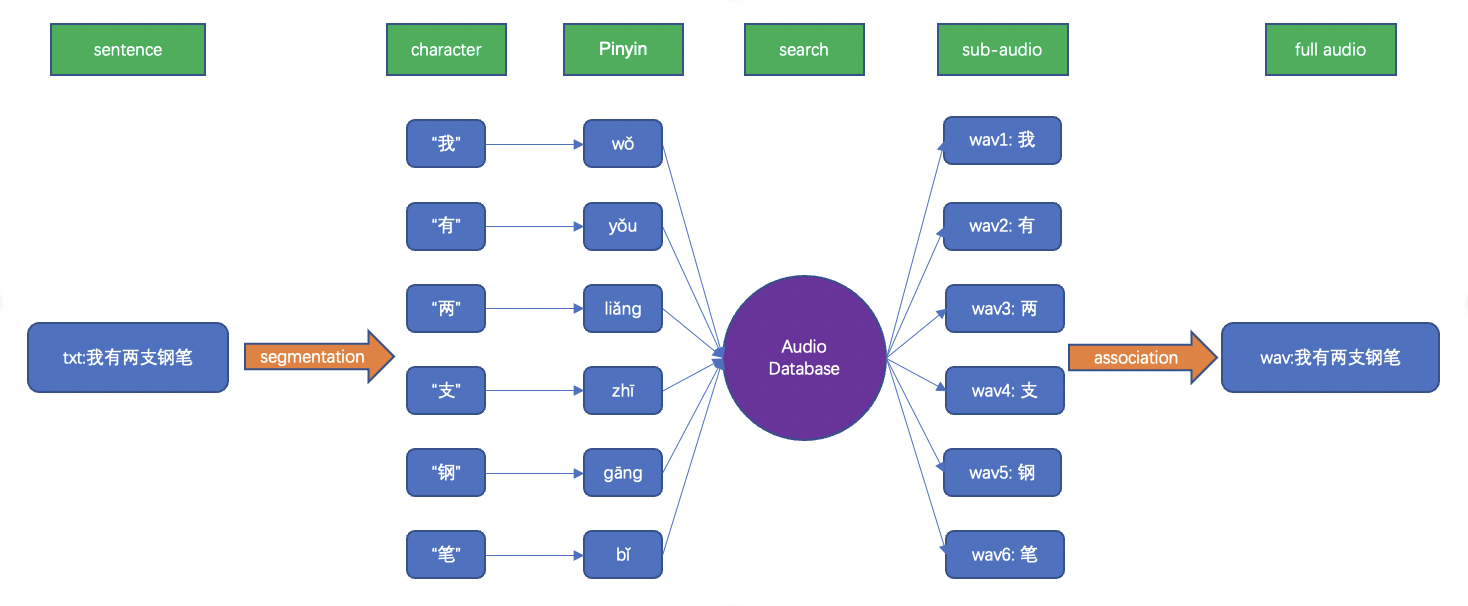}
    \caption{ The CAMP method procedure.}
    \label{f1}
\end{figure*}

Besides data augmentation, semi-supervised learning, to improve the model performance with unlabeled training data, is also  popularly applicable and successful in many scenarios. 
There are mainly two types of solutions for semi-supervised learning. The first is to take advantage of the continuity assumption that if an actual perturbation is applied to an unlabeled data, the prediction should not change significantly. Hence, minimizing the distance
between the unlabeled data and its perturbation helps to improve the model performance (\cite{laine2016temporal},  \cite{rasmus2015semi},  \cite{tarvainen2017mean}). The second is to generate pseudo labels for the unlabeled data, and then mix these pseudo-labeled data with labeled data to provide additional information for training. The validity of this method is shown in \cite{lee2013pseudo}, \cite{blum1998combining} and \cite{zhou2005tri}, even though the generated labels 
unavoidably contain incorrectness.

It is worth mentioning that the CAMP itself can be also regarded as a TTS method. However, there are important differences compared with previous ASR-TTS self-supervised learning methods (\cite{baskar2019semi}, \cite{baskar2021eat}). The main difference is that the speech data generated by CAMP is real, despite of probable concatenation of real speech. As a comparison, under the text only (TO) regime of the ASR-TTS self-supervised learning method, the audio reconstructed by the TTS module is distorted due to the lack of real audio information \cite{baskar2021eat}. Furthermore, our CAMP method not only has a more concise process, but also can control the diversity of generated audio, e.g. generating multiple corresponding audios for a fixed text, which is obviously helpful to improve the robustness of the ASR system.

\section{Methods}

\subsection{Character audio mix up (CAMP)}\label{sec:camp}
Inspired by the methods from pseudo label semi-supervised learning (\cite{lee2013pseudo}, \cite{blum1998combining}, \cite{zhou2005tri}) and mix up data augmentation (\cite{zhang2017mixup}), the insight herein is to generate pseudo audio for any given texts via mandarin character scale mixing up. 

Specifically, for each character in a mandarin sentence, we first find the matching pronunciation from the mandarin character audio database, which is denoted as $\boldsymbol{a_1},\boldsymbol{a_2},...,\boldsymbol{a_n}$. Then, we normalize the audio sequence $\boldsymbol{a_1},\boldsymbol{a_2},...,\boldsymbol{a_n}$ by 
\begin{equation}
\begin{aligned}
&E=\frac{1}{n} \sum_{i=1}^n\left\|\boldsymbol{a_i}\right\|_2, \\
&\tilde{\boldsymbol{a}}_i=\frac{\boldsymbol{a_i}}{\left\|\boldsymbol{a_i}\right\|_2}E.
\end{aligned}
\end{equation}
Here, $E$ represents the average energy of the audio sequence $\boldsymbol{a_1},\boldsymbol{a_2},...,\boldsymbol{a_n}$. By normalization, we can guarantee that these audio clips have the same energy to better match the actual scene. Finally, we splice the normalized audio sequence $\tilde{\boldsymbol{a}}_1,\tilde{\boldsymbol{a}}_2,...,\tilde{\boldsymbol{a}}_n$ to get pseudo speech audio. Following this process, we can generate audio as much as possible given enough texts. Not only that, even for a fixed text, by controlling the selection of $\boldsymbol{a_1},\boldsymbol{a_2},...,\boldsymbol{a_n}$, we can generate multiple corresponding audios. The schematic diagram of CAMP is shown in Fig.\ref{f1}.

\subsection{META-AUDIO}\label{sec:meta-audio}
For building the mandarin character audio database, we can get the mandarin character audio fragment by using a pre-trained model to force alignment. However, we can only get 
character-audio pairs by alignment, while finding enough audio for each character often requires a sufficiently large audio dataset and a powerful model to enforce alignment,  since there is a large amount of mandarin characters (more than 60k). Inspired by the well-known fact that any vector in a linear space can be represented by (the linear combination of) the basis vectors, we extract the key idea of meta audio. That is, the pronunciation of each character can be represented by the meta audio. Since mandarin is monosyllabic, i.e. each pronunciation may correspond to far more than one mandarin character, mining mandarin character with the same pronunciation can be achieved by Pinyin.

\begin{figure}
    \centering
    \includegraphics[width=.45\textwidth,height=0.25\textwidth]{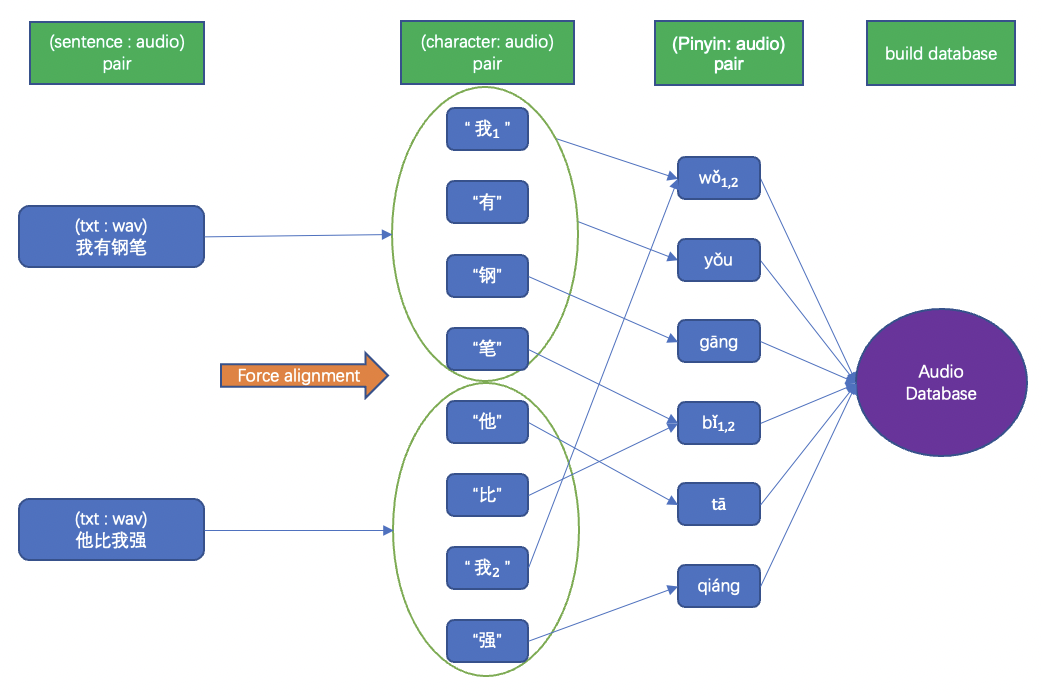}
    \caption{ The META-AUDIO method procedure.}
    \label{f2}
\end{figure}

Specifically, we can merge the character-audio pairs if the character has the same Pinyin call-out, and in the mandarin character audio database, we only save the Pinyin-audio pairs. One can view the Pinyin here as a META-AUDIO indicator. When we aim to obtain an audio example of a character, we first get the Pinyin corresponding to the character, and then index the Pinyin in the mandarin character audio database to get the audio examples. Using the META-AUDIO method, we can reduce the difficulty of database construction as well as increase the diversity of the database.  The procedure of META-AUDIO method  is shown in Fig.\ref{f2}.

\section{Experiments}

To verify the validity of our CAMP and META-AUDIO methods, we perform numerical experiments on ASR (automatic speech recognition) tasks. In this section, we present the datasets, parameters setting and results. 
\subsection{Datasets}
We construct the META-AUDIO method to build mandarin character audio database on the AISHELL-1 dataset \cite{bu2017aishell}, which is one of the most commonly used Mandarin datasets. The recorded texts in AISHELL-1 involve 11 fields including smart home, unmanned driving, and industrial production. 
We only use the training split (around 150 hours) in AISHELL-1 to build the mandarin character audio database here. 

While for generating pseudo audio by CAMP method,
in addition to the AISHELL-1 \cite{bu2017aishell} Mandarin dataset, we also conduct CAMP  on the AISHELL-3 \cite{shi2020aishell} and AIDATA-TANG\_200zh Mandarin datasets \footnote{\url{https://openslr.org/62}}. The  AISHELL-3 dataset contains 85 hours of audio and 88035 sentences. In order to split the data samples into training set and test set, we randomly select 10,000 sentences with about 10 hours of audio as the test set, and the rest as the training set. The AIDATATANG\_200zh corpus contains 200 hours of acoustic data, which is divided into the training set, validation set and test set in a ratio of 7:1:2. For the AISHELL-1, AISHELL-3 and AIDATATANG\_200zh dataset, we use the CAMP method to generate pseudo audio by texts in their training split part.
\subsection{META-AUDIO: database construction}
Using the AISHELL-1 dataset, we can build the Pinyin-audio fragment database for generating pseudo audio via META-AUDIO. First, to force alignment, we use the pre-trained architecture with a conformer encoder as well as unified two-passjoint CTC/attention \cite{watanabe2017hybrid}  decoder. The model is trained on the AISHELL-1 dataset for 240 epochs with online speech perturbations (0.9×, 1.1×) and averaged over every 20 checkpoints. Then, we use the Pinyin tool \footnote{\url{https://github.com/mozillazg/python-Pinyin}} to query the Pinyin of characters in all character-audio pairs generated by forced alignment, and convert the character-auido pairs to Pinyin-audio pairs. In the Pinyin-audio fragment database, we distinguish the tones, i.e. different tones (e.g. ’jiā’ and ’jiá’) are treated as different pronunciations. For the polyphonic Chinese characters, we select the most commonly used pronunciation (as well as the corresponding Pinyin). Finally, we save all the Pinyin-audio pairs to the mandarin character audio database.
\subsection{CAMP}
We generate the pseudo audio by CAMP on the AISHELL-1, AISHELL-3 and AIDATATANG dataset, respectively. For each character in transcriptions of the training split of AISHELL-1, AISHELL-3 and AIDATATANG, we first query the Pinyin with the Pinyin tool, then randomly choose one corresponding audio fragment in the Pinyin-audio fragment database built previously. Finally, we normalize and concatenate the corresponding audio segments for each word in  transcriptions of AISHELL-1, AISHELL-3 and AIDATATANG to 
obtain corresponding pseudo audio data. 
\subsection{ASR: experimental setup}

\begin{figure}[htb]
        \centering
        \includegraphics[width=0.3\textwidth,height=0.14\textwidth]{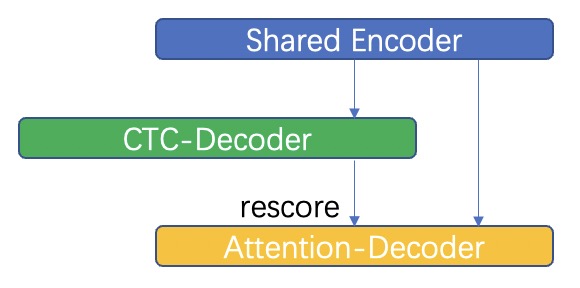}
	\caption{WeNet architecture.}
 \label{wenet_fig}
\end{figure}

The experiments are conducted using WeNet \cite{yao2021wenet}, which is based on a two-pass CTC and AED joint architecture, as is shown in Fig. \ref{wenet_fig}. In WeNet, the Shared Encoder is composed of multiple Transformers \cite{vaswani2017attention} or Conformers \cite{gulati2020conformer} in order to extract information from speech data and encode it into high-dimensional embeddings. CTC-Decoder consists of several fully-connected layers, while Attention-Decoder consists of multiple Transformer decoder layers. The whole model dynamics reads that, the input data goes through the Shared Encoder, first decoded by the CTC-Decoder as a rough selection to get initial candidates of output texts, then the outputs of encoders, candidates and their ctc-scores is passed into the Attention-Decoder to resort and produce more accurate results.
The experiments on AISHELL-1 are performed using two 24Gb memory RTX3090 GPUs with a batch size of 32 for each GPU, and we conduct experiments on AISHELL-3 and AIDATATANG with four 16Gb memory P100 GPUs with a respective batch size of 16. Therefore, all experiments actually share the same batch size through the distributed data-parallel training. The Adam optimizer with a learning rate of 0.002 is used during the training process. We train the models for 120 epochs on AISHELL-1, 70 epochs on AISHELL-3 and 100 epochs on AIDATATANG. We adopt 12 conformer layers as the Shared Encoder and 6 transformer layers as Attention-Decoder for all experiments. All embedding dimensions of transformer layers are set as 256 with 4 attention heads.


\subsection{Results}
For each dataset, we randomly select 10 hours of audio-text corpus from the corresponding training set, which is described as \emph{10h real data}, and mix it with the pseudo data obtained on this specific dataset to get the new training set.

\begin{table}[h]
\scalebox{0.7}{
\begin{tabular}{c|ccc}
\hline
           & \multicolumn{3}{c}
           {Character error rate (CER)}                              
           \\ \hline
Dataset    & \multicolumn{1}{c|}{10h real data only}      & \multicolumn{1}{c|}{10h real data + pseudo data} & full real data \\ \hline
AISHELL-1  & \multicolumn{1}{c|}{\textgreater{}60} & \multicolumn{1}{c|}{11.07}              & 4.60         \\
AISHELL-3  & \multicolumn{1}{c|}{\textgreater{}60} & \multicolumn{1}{c|}{14.74}              & 8.71         \\
AIDATATANG & \multicolumn{1}{c|}{20.38}            & \multicolumn{1}{c|}{8.26}               & 4.72         \\ \hline
\end{tabular}
}
\caption{The CER results on test sets under different training sets.}
\label{t1}
\end{table}


The results of automatic speech recognition are shown in Table \ref{t1}. Here, the shorthand \emph{‘10h real data only’} denotes that the training is only performed on 10 hours of real data, while the shorthand \emph{‘10h real data + pseudo data’} denotes that the training set consists of 10 hours of real data and obtained pseudo data. For comparison, we also train the same model with full real training data. Note that we conduct all experiments without using any language models.

From Table \ref{t1}, one can straightforwardly obtain the following observations.
If we only use 10 hours of real data as the training set, the final character error rate is rather high on all three datasets, especially for AISHELL-1 and AISHELL-3. A CER larger than 60 on test sets means that you can hardly understand any sentences. 
The terrible performance is reasonable due to the lack of training data. With the help of pseudo data, the CER on test sets decreases significantly on every dataset, which proves the effectiveness of our data augmentation methods. Compared to the results obtained on the whole training sets, our results are still competitive despite of a quite limited usage of real data.
Not only that, when we check the incorrect examples of inference texts on test audio, most of the wrong words appear similar pronunciations to the correct ones, and the whole sentences can be easily understood and corrected.

\section{Ablation studies}

To further investigate the impact of pseudo data, we also design corresponding ablation experiments. On one hand, we remove the 10 hours of real data which is randomly selected from the training set, and only use the pseudo data. The CER results on test sets are shown in Table \ref{t2}. The observation is that the real data from original training sets is really important, and its lack will lead to a great decrease of the model performance.
\begin{table}[h]
\centering
\scalebox{0.8}{
\begin{tabular}{c|cc}
\hline
           & \multicolumn{2}{c}{Character error rate (CER)} \\ \hline
Dataset    & \multicolumn{1}{c|}{10h real data + pesudo data}     & pseudo data only   \\ \hline
AISHELL-1  & \multicolumn{1}{c|}{11.07}                  & 27.38             \\
AISHELL-3  & \multicolumn{1}{c|}{14.74}                  & 30.73             \\
AIDATATANG & \multicolumn{1}{c|}{8.26}                   & 54.78             \\ \hline
\end{tabular}
}
\caption{The CER results on test sets under different training sets.}
\label{t2}
\end{table}

On the other hand, we mix up pseudo data with full real data as the training set on AISHELL-3, and get a CER result of 9.45 on the test set. Although it is still larger than the CER result of 8.71 where only the full real data is used for training (see Table \ref{t1}), the performance is much better than the CER result of 14.74 using 10 hours of real data and pseudo data. 

Combining the above two ablation experiments, we conjecture that there still exists an inevitable gap of distributions between the pseudo data and real data. Adding 10 hours of real data helps the model find an anchor of the underlying distribution (ground truth), which indicates the importance of real data. When increasing the ratio of real data in the training set, the test performance increases reasonably since the distribution for training biases towards the ground truth.

\section{Conclusion}
In this work, we propose a novel method named as META-AUDIO to build a mandarin character scale audio database. Also, we raise the CAMP procedure to generate audios from a character scale. Combining these two methods, we tend to generate pesudo speech data conveniently. Through numerical experiments on several representative datasets, one can obtain competitive CER results using limited real data and our pseudo data, which validates the great effectiveness and low (real) data dependency of our methods. For those languages (e.g. dialects) that is difficult to obtain sufficient audio data, we hope that our methods can make a great contribution.

\bibliographystyle{IEEEbib}
\bibliography{strings,refs}

\end{document}